\documentclass[%
 reprint,
 amsmath,amssymb,
 aps,
]{revtex4-1}

\usepackage{graphicx}
\usepackage{dcolumn}
\usepackage{bm}
\usepackage{comment}
\usepackage{xcolor}
\usepackage{hyperref}
\usepackage{url}

\begin{document}

\preprint{APS/123-QED}






\title{Higher-order nonequilibrium term: Effective power density quantifying evolution towards or away from local thermodynamic equilibrium}

\author{M. Hasan Barbhuiya}
\email{mhb0004@mix.wvu.edu}
\author{Paul A. Cassak}%
\author{Subash Adhikari}
\affiliation{%
Department of Physics and Astronomy and the Center for KINETIC Plasma Physics, 
West Virginia University, Morgantown, WV 26506, USA
}%

\author{Tulasi N. Parashar}
\affiliation{School of Chemical and Physical Sciences, Victoria University of Wellington, Gate 7, Kelburn Parade, Wellington, 6012, New Zealand}

\author{Haoming Liang}
\affiliation{Department of Astronomy, University of Maryland College Park, College Park, MD 20742, USA and NASA Goddard Space Flight Center, Greenbelt, MD 20771, USA
}

\author{Matthew R.~Argall}
\affiliation{Space Science Center, Institute for the Study of Earth, Oceans, and Space \\ University of New Hampshire, Durham, NH 03824, USA}


%


\date{\today}

\begin{abstract}
A common approach to assess the nature of energy conversion in a classical fluid or plasma is to compare power densities of the various possible energy conversion mechanisms.  A forefront research area is quantifying energy conversion for systems that are not in local thermodynamic equilibrium (LTE), as is common in a number of fluid and plasma systems. Here, we introduce the ``higher-order non-equilibrium term'' (HORNET) effective power density that quantifies the rate of change of departure of a phase space density from LTE. It has dimensions of power density, which allows for quantitative comparisons with standard power densities. We employ particle-in-cell simulations to calculate HORNET during two processes, namely magnetic reconnection and decaying kinetic turbulence in collisionless magnetized plasmas, that inherently produce non-LTE effects. We investigate the spatial variation of HORNET and the time evolution of its spatial average. By comparing HORNET with power densities describing changes to the internal energy (pressure dilatation, $\rm{Pi-D}$, and divergence of the vector heat flux density), we find that HORNET can be a significant fraction of these other measures (8\% and 35\% for electrons and ions, respectively, for reconnection; up to 67\% for both electrons and ions for turbulence), meaning evolution of the system towards or away from LTE can be dynamically important. Applications to numerous plasma phenomena are discussed.
\end{abstract}

\maketitle

\section{Introduction}
\label{sec:intro}

The classical laws of thermodynamics quantify the conversion of energy for processes occurring in systems in local thermodynamic equilibrium (LTE). However, many systems of interest are not in \textcolor{black}{LTE}. 
The approaches by Chapman and Enskog \cite{Chapman70}, Onsager \cite{Onsager31}, Grad \cite{Grad49} and its extensions \cite{Jou10}, and the gyrokinetic approach \cite{Howes06,Schekochihin09} have revealed many important insights \textcolor{black}{about energy conversion for systems not in LTE}
but the models assume the departures from LTE are small. Many systems are routinely far from LTE, such as plasmas in space and astrophysical settings where densities are low and temperatures are high, leading to low collisionality. This indicates there is a necessity for a non-perturbative approach to quantify non-LTE energy conversion.

One non-perturbative approach is to use the non-LTE internal (thermal) energy density evolution equation \cite{Braginskii65}, which follows directly from the Boltzmann transport equation. As will be reviewed in the next section, four physical mechanisms can cause the internal energy density to change in the comoving (Lagrangian) reference frame.  They are a) pressure dilatation, which describes compressible heating; b) incompressible heating, including deformation and shear, captured by the term called ${\rm Pi-D}$ \cite{del_sarto_pressure_2016,yang_PRE_2017,Yang17,Yang_2022_ApJ,Cassak_PiD1_2022,Cassak_PiD2_2022,Barbhuiya_PiD3_2022}; c) the divergence of the \textcolor{black}{vector} heat flux density \cite{zhou_measurements_2021,song_forcebalance_2020, Du_energy_2020_PRE}; and d) inter-species collisions. In LTE, $\mathrm{Pi-D}$, \textcolor{black}{vector} heat flux density, and the collision operator vanish, so these terms describe non-LTE energy conversion. They have been extensively used to study energy conversion in space and astrophysical plasmas, especially during magnetic reconnection and turbulence \cite{del_sarto_pressure_2016,Yang17,Chasapis18,yang_scale_2019,Pezzi19,song_forcebalance_2020,Du_energy_2020_PRE,Pezzi21,bandyopadhyay_energy_2021,zhou_measurements_2021,Yang_2022_ApJ,Cassak_PiD1_2022,Cassak_PiD2_2022,Barbhuiya_PiD3_2022}.

Another recent approach that allows for the identification of different non-LTE energy conversion mechanisms in a wide variety of plasma systems is the so-called ``field-particle correlation'' approach \cite{Klein16}. It directly identifies the energy conversion between electromagnetic fields and charged particles in a plasma as a function of the phase space density. It has been applied to many plasma systems in simulations and observations \cite{Howes17,Klein17,Chen19,Klein20,Juno21,McCubbin22,Montag22,Montag23}.

\textcolor{black}{In instances where a system deviates significantly from LTE, each of the infinite number of moments of the phase space density can evolve and become dynamically significant. As internal energy is related to only a single moment of the phase space density, a description of the evolution of the other moments can become pertinent. However, a closed set of evolution equations for each of the infinite moments is unattainable due to the closure problem. One potential strategy \textcolor{black}{to describing every moment is to employ} 
the Boltzmann (or Vlasov) transport equation for the phase space density itself, \textcolor{black}{but this} is challenging since the phase space density is a six-dimensional (plus time) field.  Notably, recent research \cite{Cassak_FirstLaw_2023} has demonstrated that the evolution of all moments associated with a non-LTE system can be described by the relative entropy (that is a three-dimensional plus time field).} 
It is \textcolor{black}{a first-principle derivation from} 
the Boltzmann equation using an entropic approach \cite{Liang19}. 

Mechanisms that can change the internal energy (pressure dilatation, ${\rm Pi-D}$, and divergence of \textcolor{black}{vector} heat flux density), are typically quantified as power densities, {\it i.e.,} rates at which the energy density changes. \textcolor{black}{In addition, $\mathbf{J}_\sigma \cdot \mathbf{E}$}
\textcolor{black}{describes} the \textcolor{black}{rate of} conversion of energy \textcolor{black}{density} between electric fields and charged particles \textcolor{black}{of species $\sigma$,} where \textcolor{black}{$\mathbf{J}_\sigma$ is the current density of 
particle species $\sigma$} and $\mathbf{E}$ is the electric field. \textcolor{black}{Previous studies \cite{Yang17, Pezzi19} have demonstrated that $\mathbf{J}_\sigma \cdot \mathbf{E}$ does not directly result in a change in internal energy and can only 
change the bulk kinetic energy.
While questions persist regarding the precise mechanism through which electric fields can induce changes in internal energy, we provide only a cursory treatment of this energy conversion channel because it does not directly go to internal energy and has been compared to other internal energy measures previously.}

Here, we use the result of Ref.~\cite{Cassak_FirstLaw_2023} to define \textcolor{black}{a quantity $P_{\sigma,{\rm HORNET}}$ describing the rate of change of the departure of a phase space density from LTE that has dimensions of power density, where HORNET stands for ``higher-order non-equilibrium term.''}
\textcolor{black}{
We call it an effective power density because it is not a power density in the sense that it does not describe a rate of change of energy density.} 
In this study, we calculate $P_{\sigma,{\rm HORNET}}$ in particle-in-cell (PIC) simulations of magnetic reconnection and \textcolor{black}{decaying plasma} turbulence. By locally comparing $P_{\sigma, {\rm HORNET}}$ to other power densities, we can ascertain the relative prevalence of the 
\textcolor{black}{evolution of a phase space density towards or away from LTE
compared to the energy conversion taking place.} We discover that the spatially averaged HORNET power density can be a substantial 
\textcolor{black}{percentage} of the net power density changing the internal energy at the times of interest; for \textcolor{black}{a simulation of} reconnection, we find that this \textcolor{black}{percentage} in the electron diffusion region is 8\% and 35\% for electrons and ions, respectively; in \textcolor{black}{a simulation of} turbulence, this \textcolor{black}{percentage} for the entire domain \textcolor{black}{is} 
67\% for both electrons and ions. We expect $P_{\sigma,{\rm HORNET}}$ will be useful for many non-LTE processes in plasma physics and other fields of science.

The layout of this paper is as follows.  Sec.~\ref{sec:theory} contains a review of the relevant results from Ref.~\cite{Cassak_FirstLaw_2023} and introduces $P_{\sigma,{\rm HORNET}}$. Sec.~\ref{sec:simulation} describes the numerical simulation setup. Secs.~\ref{subsec:reconn} and~\ref{subsec:turbu} give the simulation results for a simulation of magnetic reconnection and decaying plasma turbulence, respectively.  Sec.~\ref{sec:conclusion} summarizes the results and discusses applications and implications.

\section{Theory}
\label{sec:theory}

\subsection{Review}

The \textcolor{black}{non-relativistic} Boltzmann transport equation \cite{Boltzmann1872} for species $\sigma$ is given by
\begin{equation}
    \frac{\partial f_\sigma}{\partial t} + {\bf v} \cdot \nabla f_\sigma + \frac{{\bf F}_\sigma}{m_\sigma} \cdot \nabla_v f_\sigma = C[f],
    \label{eq:boltzeqn}
\end{equation}
where $f_\sigma$ is the phase space density, ${\bf F}_\sigma$ is the net body force, $m_\sigma$ is the mass of the constituents, ${\bf r}$, ${\bf v}$, and $t$ are the position-space, velocity-space, and time coordinates of phase space, $\nabla$ and $\nabla_v$ are the position- and velocity-space gradient operators, \textcolor{black}{respectively,} and $C[f]$ is a collision operator that may be a sum of intra-species and inter-species collision operators given by $C[f] = C_{\sigma}[f_\sigma] + \sum_{\sigma^\prime \neq \sigma} C_{\sigma \sigma^\prime}[f_\sigma,f_{\sigma^\prime}]$. Multiplying Eq.~(\ref{eq:boltzeqn}) by $(1/2) m_\sigma v_\sigma^{\prime 2}$ and integrating over all velocity space, where ${\bf v}_\sigma^\prime = {\bf v} - {\bf u}_\sigma$ is the random velocity, ${\bf u}_\sigma = (1/n_\sigma) \int f_\sigma {\bf v} d^3v$ is the bulk flow velocity and $n_\sigma = \int f_\sigma d^3v$ is the number density, one obtains the time evolution equation of internal energy density $\bar{\mathcal{E}}_{\sigma,{\rm int}} = \int (1/2) m_\sigma v_\sigma^{\prime 2} f_\sigma d^3v$ given by \cite{Braginskii65}
\begin{eqnarray}
    \frac{\partial \bar{\mathcal{E}}_{\sigma,{\rm int}}}{\partial t} & + & \boldsymbol{\nabla} \cdot (\bar{\mathcal{E}}_{\sigma,{\rm int}} {\bf u}_\sigma) = \nonumber \\  & - & ({\bf P}_\sigma \cdot \boldsymbol{\nabla}) \cdot {\bf u}_\sigma - \boldsymbol{\nabla} \cdot {\bf q}_\sigma + n_\sigma \dot{Q}_{\sigma,{\rm coll,inter}}, \label{eq:IntEnerEq}
\end{eqnarray}
where ${\bf P}_{\sigma}=n_\sigma k_B {\bf T}_{\sigma}$ is the pressure tensor, ${\bf T}_\sigma$ is the temperature tensor with elements $T_{\sigma,jk} = m_\sigma/(n_\sigma k_B) \int v_{\sigma j}^\prime v_{\sigma k}^\prime f_\sigma d^3v$, $k_B$ is Boltzmann's constant, ${\bf q}_\sigma = \int (1/2) m_\sigma v_\sigma^{\prime 2} {\bf v}_\sigma^\prime f_\sigma d^3v$ is the vector heat flux density, and $\dot{Q}_{\sigma,{\rm coll,inter}} = (1/n_\sigma) \int (1/2) m_\sigma v_\sigma^{\prime 2} \sum_{\sigma^\prime \neq \sigma} C_{\sigma \sigma^\prime}[f_\sigma,f_{\sigma^\prime}] d^3v$ is the volumetric heating rate per particle due to inter-species collisions. The pressure-strain interaction is defined as $- ({\bf P}_\sigma \cdot \boldsymbol{\nabla}) \cdot {\bf u}_\sigma$ \cite{del_sarto_pressure_2016,Yang_2022_ApJ,Cassak_PiD1_2022,Yang17,yang_PRE_2017}. Note, $\bar{\mathcal{E}}_{\sigma,{\rm int}} = 3 {\cal P}_\sigma / 2 = (3/2) n_\sigma k_B \mathcal{T}_\sigma$, where we call ${\cal P}_\sigma = {\rm tr} ({\bf P}_\sigma)/3$ the effective pressure and ${\cal T}_\sigma = {\rm tr}({\bf T}_\sigma)/3$ the effective temperature, and tr is the trace. \textcolor{black}{The lack of the ${\bf J}_\sigma \cdot {\bf E}$ term in this equation is used as evidence that the term does not directly change the internal energy density.}

It is advantageous to write Eq.~(\ref{eq:IntEnerEq}) in the comoving (Lagrangian) reference frame using the convective derivative $d/dt = \partial/\partial t + {\bf u}_\sigma \cdot \nabla$, which gives
\cite{Braginskii65}
\begin{equation}
\frac{d{\cal E}_{\sigma,{\rm int}}}{dt} = \frac{- {\cal P}_\sigma (\boldsymbol{\nabla} \cdot {\bf u}_\sigma) -\Pi_{\sigma,jk} {\cal D}_{\sigma,jk} - \boldsymbol{\nabla} \cdot {\bf q}_\sigma}{n_\sigma} + \dot{Q}_{\sigma,{\rm coll,inter}}, \label{eq:intenergyevolve}
\end{equation}
where ${\cal E}_{\sigma,{\rm int}} = (3/2) k_B {\cal T}_\sigma$ (without the overbar) is the internal energy per particle. In writing Eq.~(\ref{eq:intenergyevolve}), we decompose the pressure-strain interaction into a sum of pressure dilatation $-{\cal P}_\sigma (\boldsymbol{\nabla} \cdot {\bf u}_\sigma)$ and ${\rm Pi-D}_\sigma = -\Pi_{\sigma,jk} {\cal D}_{\sigma,jk}$ that describe compressible and incompressible energy conversion between bulk kinetic energy and internal energy, respectively, and we use the Einstein summation convention for repeated dummy indices. Here, $\Pi_{\sigma,jk} = P_{\sigma,jk} - {\cal P}_\sigma \delta_{jk}$ is the $jk$ element of the deviatoric pressure tensor, ${\cal D}_{\sigma,jk} = (1/2) \left( \partial u_{\sigma j}/\partial r_k + \partial u_{\sigma k}/\partial r_j \right) - (1/3) \delta_{jk} (\boldsymbol{\nabla} \cdot {\bf u}_\sigma)$ is the $jk$ element of the traceless symmetric strain rate tensor, and $\delta_{jk}$ is the Kronecker delta \cite{Yang17}. 

In Ref.~\cite{Cassak_FirstLaw_2023}, it was shown \textcolor{black}{using an entropic approach that for systems not in} 
LTE with a fixed total number of particles $N_\sigma$ of each species, \textcolor{black}{the} so-called first law of kinetic theory \textcolor{black}{is} given by

\begin{equation}
\frac{d{\cal W}_\sigma}{dt} + \frac{d{\cal E}_{\sigma,{\rm gen}}}{dt} = \frac{d{\cal Q}_{\sigma,{\rm gen}}}{dt} + \dot{{\cal Q}}_{\sigma,\rm coll}, \label{eq:firstlawfinal}
\end{equation}
where $d{\cal E}_{\sigma,{\rm gen}}$ and $d{\cal Q}_{\sigma,{\rm gen}}$ are increments in the so-called generalized energy per particle and generalized heat per particle, respectively, and
\begin{subequations}
\begin{eqnarray}
\frac{d{\cal W}_\sigma}{dt} & = & {\cal P}_\sigma \frac{d(1/n_\sigma)}{dt}, \label{eq:work} \\ \frac{d{\cal E}_{\sigma,{\rm gen}}}{dt} & = & \frac{d{\cal E}_{\sigma,{\rm int}}}{dt} + \frac{d{\cal E}_{\sigma,{\rm rel}}}{dt}, \label{eq:genenergy} \\ \frac{d{\cal Q}_{\sigma,{\rm gen}}}{dt} & = &  \frac{d{\cal Q}_{\sigma}}{dt} + \frac{d{\cal Q}_{\sigma,{\rm rel}}}{dt}, \label{eq:genheat} \\ \dot{{\cal Q}}_{\sigma,{\rm coll}} & = & -\frac{k_B \mathcal{T}_\sigma}{n_\sigma} \int C[f] \ln \left( \frac{f_\sigma \Delta^3r_\sigma \Delta^3 v_\sigma}{N_\sigma} \right) d^3v.  \label{eq:qdotcoll}
\end{eqnarray}
\end{subequations}
Equation~(\ref{eq:work}) describes the rate of change of compressional work per particle $d{\cal W}_\sigma /dt$ done by the system. The two terms in Eq.~(\ref{eq:genenergy}) describe the rate of change of internal energy per particle $d{\cal E}_{\sigma,{\rm int}}/dt$ and the rate of change of \textcolor{black}{effective} energy per particle \textcolor{black}{(the relative energy per particle) describing the local evolution towards or away from LTE $d{\cal E}_{\sigma,{\rm rel}}/dt$, respectively.} 
The two terms in Eq.~(\ref{eq:genheat}) describe the heating rate per particle associated with changes to the internal energy $d{\cal Q}_{\sigma}/dt = (-\nabla \cdot {\bf q}_\sigma - \Pi_{\sigma,jk} {\cal D}_{\sigma,jk}) / n_\sigma$ [see Eq.~(\ref{eq:intenergyevolve})] and 
\textcolor{black}{an effective rate of heat per particle associated with the local phase space density not being in LTE $d{\cal Q}_{\sigma,{\rm rel}}/dt$, respectively.} Equation~(\ref{eq:qdotcoll}) describes the rate of energy change per particle associated with collisions. In Eq.~(\ref{eq:qdotcoll}), $\Delta^3r_\sigma \Delta^3v_\sigma$ is the six-dimensional phase space volume element \cite{Liang19,Liang20,Argall22}.

Quantitatively, the two relative terms  
\textcolor{black}{associated with the departure of the phase space density from LTE are} given by
\begin{subequations}
\begin{eqnarray}
    \frac{d{\cal E}_{\sigma,{\rm rel}}}{dt} & = & {\cal T}_\sigma \frac{d(s_{\sigma v,{\rm rel}}/n_\sigma)}{dt}, \label{eq:msigreldef} \\
    \frac{d{\cal Q}_{\sigma,{\rm rel}}}{dt} & = & -{\cal T}_\sigma \frac{(\nabla \cdot \bm{\mathcal{J}}_{\sigma,{\rm th}})_{{\rm rel}}}{n_\sigma}, \label{eq:ddtqreldef}
\end{eqnarray}
\end{subequations}
where the relative entropy per particle \cite{Grad65} and the thermal relative entropy density flux divergence per particle \cite{Cassak_FirstLaw_2023} are defined as 
\begin{subequations}
    \begin{eqnarray}
    \frac{s_{\sigma v,{\rm rel}}}{n_\sigma} & = & -\frac{k_B}{n_\sigma} \int f_\sigma \ln \left(\frac{f_\sigma}{f_{\sigma M}}\right) d^3v, \label{eq:relentr} \\
    \frac{(\boldsymbol{\nabla} \cdot \bm{\mathcal{J}}_{\sigma,{\rm th}})_{{\rm rel}}}{n_\sigma} & = &- \frac{k_{B}}{n_\sigma} \int \left[ \frac{}{} \boldsymbol{\nabla} \cdot ({\bf v}_\sigma^\prime f_\sigma)\right] \ln\left(\frac{f_\sigma}{f_{\sigma M}}\right) d^{3} v.  \label{eq:divjthrel}
    \end{eqnarray}
\end{subequations}
Here, $f_{\sigma M}$ is the ``Maxwellianized'' phase space density associated with $f_\sigma$ given by \cite{Grad65}
\begin{equation}
f_{\sigma M} = n_\sigma \left(\frac{m_\sigma}{2\pi k_B {\cal T}_\sigma}\right)^{3/2} e^{{-m_\sigma({\bf v} - {\bf u}_\sigma)^2/2 k_B {\cal T}_\sigma}}, \label{eq:maxwellian}
\end{equation}
with $n_\sigma$, ${\bf u}_\sigma$, and ${\cal T}_\sigma$ obtained from $f_\sigma$. 

\textcolor{black}{Because} $s_{\sigma v,{\rm rel}}/n_\sigma$ identically vanishes if $f_\sigma = f_{\sigma M}$ \cite{Grad65}, so  relative entropy per particle contains information about the non-Maxwellianity of a local phase space density, and its \textcolor{black}{Lagrangian} time derivative describes how the Maxwellianity locally changes in time.

\subsection{The Higher Order Non-Equilibrium Term Power Density $P_{\sigma,{\rm HORNET}}$}
\label{subsec:TheoNew}

Here, we construct a quantity describing 
\textcolor{black}{the evolution of a phase space density towards or away from LTE} with dimensions of power density to be on the same footing as the pressure-strain interaction \textcolor{black}{and} the divergence of the \textcolor{black}{vector} heat flux density. 
We note from Eq.~(\ref{eq:intenergyevolve}) that these quantities give the time rate of change in the Lagrangian frame of the internal energy per particle scaled by the number density $n_\sigma$. We, therefore, argue the analogous \textcolor{black}{effective} power density for \textcolor{black}{describing the evolution of a phase space density towards or away from LTE} 
is $-n_\sigma d{\cal E}_{\sigma,{\rm rel}}/dt$.  We call this the HORNET \textcolor{black}{effective} power density $P_{\sigma,{\rm HORNET}}$:
\begin{equation}
    P_{\sigma, {\rm HORNET}} = -n_\sigma \frac{d{\cal E}_{\sigma,{\rm rel}}}{dt} = -n_\sigma \mathcal{T}_\sigma \frac{d}{dt} \left(\frac{s_{\sigma v,{\rm rel}}}{n_\sigma}\right).
    \label{eq:pwrHORNET}
\end{equation}
Since a Maxwellian phase space density is the maximum entropy state for a fixed number of particles and internal energy \cite{Boltzmann1872}, a positive $P_{\sigma, {\rm HORNET}}$ indicates that the phase space density is evolving away from Maxwellianity in the next instant of time in the Lagrangian reference frame and a negative $P_{\sigma, {\rm HORNET}}$ indicates that the phase space density is evolving towards Maxwellianity in the next instant of time, \textit{i.e.,} it thermalizes. The magnitude quantifies \textcolor{black}{the rate that the shape of the phase space density changes scaled by the effective pressure ${\cal P}_\sigma$.} 

\textcolor{black}{To isolate the portion of Eq.~(\ref{eq:firstlawfinal}) associated with the departure of the phase space density from LTE,} we derive an equation for the time evolution of $d{\cal E}_{\sigma,{\rm rel}}$ (see also Ref.~\cite{Eu95}). Substituting Eqs.~(\ref{eq:work}) - (\ref{eq:genheat}) into Eq.~(\ref{eq:firstlawfinal}) and using Eq.~(\ref{eq:intenergyevolve}) gives
\begin{equation}
\frac{d {\cal E}_{\sigma,{\rm rel}}}{dt} = \frac{d {\cal Q}_{\sigma,{\rm rel}}}{dt} + \dot{Q}_{\sigma,{\rm coll}} - \dot{Q}_{\sigma,{\rm coll,inter}}. \label{eq:dsvrelondt}
\end{equation}
Assuming collisions conserve particle number and are elastic, the collision operator satisfies the relations $\int C_{\sigma} d^3v = \int C_{\sigma \sigma^\prime} d^3v = 0$ from conservation of particle number, $\int m_\sigma {\bf v} C_{\sigma} d^3v = 0$ from conservation of momentum for intra-species collisions, and $\int (1/2) m_\sigma v^2 C_{\sigma} d^3v = 0$ from conservation of energy for intra-species collisions. A brief derivation reveals that $\dot{Q}_{\sigma,{\rm coll,inter}}$ is equivalent to
\begin{equation}
    \dot{Q}_{\sigma,{\rm coll,inter}} = -\frac{k_B \mathcal{T}_\sigma}{n_\sigma} \int C[f] \ln \left( \frac{f_{\sigma M} \Delta^3r_\sigma \Delta^3 v_\sigma}{N_\sigma} \right) d^3v,  \label{eq:qdotcollinter}
\end{equation}
where $f_{\sigma M}$ is defined in Eq.~(\ref{eq:maxwellian}).  Then, Eq.~(\ref{eq:dsvrelondt}) becomes
\begin{equation}
\frac{d {\cal E}_{\sigma,{\rm rel}}}{dt} = \frac{d {\cal Q}_{\sigma,{\rm rel}}}{dt} + \dot{Q}_{\sigma,{\rm coll,rel}}, \label{eq:dsvrelondt2}
\end{equation}
where
\begin{equation}
    \dot{Q}_{\sigma,{\rm coll,rel}} = -\frac{k_B \mathcal{T}_\sigma}{n_\sigma} \int C[f] \ln \left( \frac{f_{\sigma}}{f_{\sigma M}} \right) d^3v. \label{eq:qdotcollrel}
\end{equation}
Eq.~(\ref{eq:dsvrelondt2}) is a special case of Eq.~(3.13) from Ref.~\cite{Eu95}. 
It may seem as though Eq.~(\ref{eq:dsvrelondt2}) implies that the relative terms are decoupled from the terms related to the internal energy [see Eq.~(\ref{eq:intenergyevolve})], but this is not the case \cite{Cassak_FirstLaw_2023} because the thermodynamic heat $d{\cal Q}_{\sigma}/dt$ terms depend on 
non-Maxwellian features of the phase space density, namely ${\bm \Pi}_\sigma$ and ${\bf q}_\sigma$, and their evolution is \textcolor{black}{included in} $d{\cal E}_{\sigma,{\rm rel}}/dt$.  Thus, Eq.~(\ref{eq:dsvrelondt2}) shows that the HORNET \textcolor{black}{effective} power density describes 
\textcolor{black}{evolution towards or away from LTE} that occurs as a result of relative heat (non-LTE features that change the 
\textcolor{black}{shape} of the phase space density) and/or collisions. 

\textcolor{black}{To better show that the relative entropy is associated with changes in shape of the phase space density, we rewrite $s_{\sigma v,{\rm rel}}/n_\sigma$ from Eq.~(\ref{eq:relentr}) as 
   \begin{eqnarray}
        \frac{s_{\sigma v,{\rm rel}}}{n_\sigma} & = & -k_B \int f^\prime_\sigma \ln \left(\frac{f^\prime_\sigma}{f^\prime_{\sigma M}}\right) d^3v, \label{eq:relentr2}
    \end{eqnarray}
where we define the distribution function as $f_\sigma^\prime = f_\sigma / n_\sigma$. Here, $f_\sigma^\prime$ is normalized to 1, while $f_\sigma$ is normalized to $n_\sigma$. By definition, $f_\sigma^\prime$ and $f_{\sigma M}^\prime$ are independent of $n_\sigma$, {\it i.e.,} they are both unchanged if $f_\sigma$ changes in such a way that its shape does not change even if the density does change. Thus $s_{\sigma v,{\rm rel}}/n_\sigma$ is independent of changes solely in $n_\sigma$. Since the HORNET effective power density is proportional to $d(s_{\sigma v,{\rm rel}}/n_\sigma)/dt$, it is zero for any process that only changes the density.}


\section{Numerical Simulations} 
\label{sec:simulation}

Numerical simulations are performed using {\tt p3d} \cite{zeiler:2002}, a massively parallel PIC code. All simulations are 3D in velocity-space, and 2.5D in position-space, \textit{i.e.}, vectors have three components and there is one invariant spatial dimension. Macro-particles are stepped forward using a relativistic Boris particle stepper \cite{birdsall91a} and electromagnetic fields are stepped using the trapezoidal leapfrog method \cite{guzdar93a}. To enforce Poisson's equation, {\tt p3d} utilizes the multigrid method \cite{Trottenberg00} to clean the electric field. We employ periodic boundary conditions in both spatial directions for all simulations. 

Simulated quantities are presented in normalized units. Lengths are normalized to the ion inertial scale $d_{i0} = c/\omega_{pi0}$, where $c$ is the speed of light, $\omega_{pi0} = (4 \pi n_0 q_i^2 /m_i)^{1/2}$ is the ion plasma frequency based on a reference number density $n_0$ and $q_i$ and $m_i$ are the ion charge and mass, respectively. Magnetic fields are normalized to a reference magnetic field $B_0$. Velocities are normalized to the Alfv\'en speed $c_{A0} = B_0/(4 \pi m_i n_0)^{1/2}$.  Times are normalized to the inverse ion cyclotron frequency $\Omega_{ci0}^{-1}= (q_i B_{0} / m_{i} c)^{-1}$. Temperatures are normalized to $m_i c_{A0}^2/k_B$, vector heat flux densities are in units of $(B_0^2/4 \pi) c_{A0}$, and power densities are in units of $(B_0^2/4 \pi) \Omega_{ci0}$.

Unrealistic values of the speed of light $c=15$ and electron-to-ion mass ratio $m_e/m_i = 0.04$ are employed for numerical expedience, but we expect these values do not qualitatively change the results presented here. To calculate kinetic entropy, we employ the implementation from Ref.~\cite{Argall22} and optimize the velocity-space grid scale by checking the agreement between the simulated kinetic entropy density $s_\sigma$ for various $\Delta v_\sigma$ (where $\sigma=e$ or $i$ for electrons or ions) and the theoretical value at $t=0$ \cite{Liang20b}. 

\textcolor{black}{To reduce the effect of PIC noise in the 2D plots that follow,
we implement} \textcolor{black}{a recursive smoothing procedure on the raw simulation data. We smooth over a width of four cells four times, followed by taking spatial or temporal derivatives, and then recursively smooth over four cells four times again. We determine the appropriateness of this smoothing procedure by testing various smoothing widths and iterations. Our selection is based on evaluating how spatial variations 
change with different smoothing parameters, ensuring that clearer, sharper structures are observed in 2D plots and 1D cuts through the domain without the occurrence of oversmoothing (not shown). 
We ascertain that using six or more iterations results in oversmoothing \textcolor{black}{for the simulations in the present study,} 
particularly in the vicinity of the X-line at $y = y_0$ \textcolor{black}{in the reconnection simulation.}}


\subsection{Magnetic Reconnection Simulations}
\label{subsec:ReconSim}

The reconnection simulation system size is $L_x \times L_y = 12.8 \times 6.4$ with $1024 \times 512$ grid cells that are initialized with 6,400 weighted particles per grid (PPG). \textcolor{black}{The use of weighted particles improves load imbalance 
by making regions of high density have a similar number of particles for the processors to advance in time as in low density regions.} 
Our choice of the simulation system size ensures that boundary effects do not interfere with the physics near the electron diffusion region (EDR) where we focus our interest. 

We initialize the simulation using the standard double tanh magnetic field profile $B_x(y) = \tanh{\left[(y-L_y/4)/w_0\right]} - \tanh{\left[(y-3L_y/4)/w_0\right]} -1$, with no initial out-of-plane (guide) magnetic field, where $w_0=0.5$ is the initial half-thickness of the current sheet. Therefore, the reference magnetic field $B_0$ is the initial asymptotic magnetic field strength. At initialization, the electron and ion density profiles are $n(y) = \{1/[2(T_{e}+T_{i})] \} \{ {\rm sech}^{2}\left[(y-L_y/4)/w_0\right] + {\rm sech}^{2}\left[(y-3L_y/4)/w_0\right] \} + n_{up}$, where the initial asymptotic upstream plasma density $n_{up}$ has the value of 0.2, and the difference between the peak current sheet number density and the upstream number density is the reference density $n_0$. The electron temperature $T_{e}$ and ion temperature $T_{i}$ are uniform and $T_{e}=1/12$, $T_{i}=5T_{e}$ at initialization, \textcolor{black}{which is inspired by \textcolor{black}{the temperature ratio} in Earth's magnetotail.} Both species are loaded as drifting Maxwellian velocity distribution functions with electron and ion out-of-plane drift speeds satisfying $u_{e,z}/T_e = -u_{i,z}/T_i$ initially.  We seed an X- and O-line pair in each of the two current sheets to initiate reconnection using a magnetic field perturbation of the form $\delta B_x =  -B_{pert} \sin \left(2 \pi x/L_x \right) \sin\left(4\pi y/L_y \right)$ and $\delta B_y =  B_{pert} \left[L_y/(2 L_x)\right] \cos\left(2 \pi x/L_x\right) \left[1-\cos\left(4\pi y/L_y\right)\right]$ with $B_{pert} = 0.05$. \textcolor{black}{The rationale behind selecting this perturbation amplitude is that it reduces the computational cost of the simulation by quickening the transition of the reconnecting system into the non-linear growth phase which is when substantial deviations from Maxwellian behavior are expected to occur. This initial perturbation amplitude is not expected to modify the nonlinear phase of reconnection.}

The electric field is cleaned every 10 particle time steps \textcolor{black}{to ensure Poisson's equation is satisfied and improve energy conservation}.  The smallest length scale of the system is the electron Debye length $\lambda_{De}=0.0176$; we choose a grid-length $\Delta=0.0125$. The smallest time scale of the system is the inverse of electron plasma frequency $\omega^{-1}_{pe}=0.012$ and we choose the particle time-step $\Delta t=0.001$, with the field time step of half of this. These choices result in excellent total energy conservation, which we need due to the small signals of energy conversion as we discuss in Sec.~\ref{subsec:reconn}; we see an increase in total energy by only 0.032\% by $t=16$. The plots we show are from the lower current sheet at time $t = 13$, during \textcolor{black}{non-linear growth} when the rate of reconnection is increasing most rapidly in time.

For the calculation of entropies, the range of the velocity space in each direction is restricted to $[-v_{lim,\sigma},v_{lim,\sigma}]$ \cite{Liang19}, where $v_{lim,\sigma}$ is chosen to be less than $c$ but contain all the particles.
We use $v_{lim,e}=14.2$, and $v_{lim,i}=7.8$. Each velocity space direction is discretized using $n_{bin,\sigma}$ bins, where $n_{bin,e}=26$, and $n_{bin,i}=34$. This gives the optimal velocity space grid scales $\Delta v_{\sigma}$, which are $\Delta v_e=1.092$ and $\Delta v_i=0.459$ \textcolor{black}{using the procedure described in Refs.~\cite{Liang19,Liang20,Liang20b,Argall22}.} The agreement with the theoretical values at $t = 0$ are within $\pm 0.7\%$ and $\pm 1.5\%$ in the far upstream region for electrons and ions, respectively, and within $\pm 2\%$ and $\pm 3\%$ at the center of the current sheet for electrons and ions, respectively. \textcolor{black}{By $t=16$, the total entropy 
is conserved to 2.08\% and 2.29\% for electrons and ions, respectively.}

In Appendices \ref{sec:appendixe} and \ref{sec:appendixq}, we plot electron reduced phase space densities (with one velocity dimension integrated out). The data are from a different simulation with the same parameters used in this study, except for PPG=25,600 instead of 6,400 \cite{Cassak_FirstLaw_2023}. We use a spatial domain of size $0.0625 \times 0.0625$ centered around the location for which phase space densities are plotted and bin the particles with a velocity space bin of size 0.1 in all three velocity directions. Reduced phase space densities are in units of $n_0/c^2_{A0}$.

\subsection{Turbulence Simulations}
\label{subsec:TurbSim}
The turbulence simulation uses a periodic domain of size $L_x \times L_y = 37.3912 \times 37.3912$ with 1024 x 1024 grid cells initialized with 6,400 unweighted PPG. 
\textcolor{black}{Using weighted particles is unnecessary for this simulation because the density is initially uniform.}
There is an initial uniform magnetic field that has a strength given by the reference value $B_0=1$ and points along $\hat{z}$. The velocity and magnetic field fluctuations are Alfv\`enic with random phase and are excited in a band of wave numbers $k \in [2,4] \times 2 \pi/L_x$ with a flat spectrum. The initial root mean square fluctuation amplitude for both velocity and magnetic field are set to 0.25. 
The initial cross helicity is very small. These initial conditions are analogous to those in a larger domain in Ref.~\cite{parashar_2018_ApJL}.

The initial number density is uniform with a value given by the reference density $n_0=1$. The electron temperature $T_{e}$ and ion temperature $T_{i}$ are initially uniform and equal to $0.3$. Both electrons and ions are initially drifting Maxwellian velocity distribution functions at the initial temperature and drifting with the local bulk flow speed. The electric field is cleaned every 40 particle time steps. \textcolor{black}{The cadence of electric field cleaning is determined by performing four simulations with the cadence set to 20, 40, 50, and 100 particle time steps, respectively, and selecting the value that produced the best energy conservation.} The smallest length scale of the system is the electron Debye length $\lambda_{De} \simeq 0.03651$ and we choose a grid-length $\Delta=\lambda_{De}$. The smallest time scale of the system is the inverse of electron plasma frequency $\omega^{-1}_{pe} \simeq 0.0133$; we choose the particle time-step to be $\Delta t=0.005$ with the field time step $\Delta t/3$. This results in excellent total energy conservation of 0.013\% by the final time of $t=50$. The non-linear time, the time scale over which non-linear interactions between different wave modes become significant, is $\tau_{nl} = L_x/[2 \pi (\delta b_{rms}^2 + \delta v_{rms}^2)^{1/2}] \simeq 17$. 

For calculating entropies, we obtain an optimal electron velocity space grid scale of $\Delta v_e=2.028$ using $v_{lim,e}=14.2$, and $n_{bin,e}=14$. At $t = 0$, the choice for $\Delta v_e$ gives agreement with the theoretical values within $\pm 0.16\%$ on average along horizontal and vertical cuts through the center of the simulation domain. For ions, using $v_{lim,i}=5.7$, and $n_{bin,i}=50$, we obtain $\Delta v_i=0.228$ which gives agreement within $\pm 0.12\%$ of theoretical values on average along the same cuts at $t = 0$. \textcolor{black}{By the final time of $t=50$, the total electron and ion entropies 
are conserved to 0.28\% and 0.55\%, respectively.}

\subsection{Relative Entropy Implementation}
\label{subsec:RelEntrSim}

In a previous study \cite{Cassak_FirstLaw_2023}, the calculation of $(d/dt)(s_{\sigma v,{\rm rel}}/n_\sigma)$ was carried out by post-processing output data of $s_\sigma$ and other fluid moments that had been rounded. For the present study, we implement the quantities natively within {\tt p3d} to reduce rounding errors. We find the overall 2D structures in $(d/dt)(s_{\sigma v,{\rm rel}}/n_\sigma)$ are similar when calculated in the two different ways (not shown).

To assess the importance of numerical error, we note that $f_\sigma$ at $t = 0$ in both the reconnection and turbulence simulations are loaded as drifting Maxwellians. Theoretically, from Eq.~(\ref{eq:relentr}), we would expect $s_{\sigma v,{\rm rel}}/n_\sigma$ to vanish initially everywhere in the simulation domain. However, these initial Maxwellians have PIC noise due to a finite number of particles per grid, so $s_{\sigma v,{\rm rel}}/n_\sigma$ is non-zero. We confirm that the error of non-zero $s_{\sigma v,{\rm rel}}/n_\sigma$ values in our reconnection simulation is due to finite PPG by comparing electron data in simulations with different PPG. For the reconnection simulation with PPG = 6,400 used in this manuscript, the average value of $s_{e v,{\rm rel}}/n_e$ over the whole domain at $t = 0$ is $\simeq -0.013$. In a test simulation with PPG = 400, we find the average value of $s_{e v,{\rm rel}}/n_e$ at $t= 0$ is $\simeq -0.08$. Noise in PIC simulations scales with $1/\sqrt{\rm PPG}$, so we expect this decrease in PPG by a factor of 16 should increase the noise by a factor of 4. This is close to the factor by which we observe the decrease, suggesting that the values of relative entropy per particle calculated within our code is a result of using finite PPG. This provides evidence that our entropy implementation is valid at $t = 0$. \textcolor{black}{This lays out an approach for future studies of relative entropy with PIC codes by measuring the desired quantity with a particular value of PPG and then increasing PPG to reduce the noise according to $1/\sqrt{{\rm PPG}}$ until the signal is higher than the noise.}

\section{Results: Magnetic Reconnection} \label{subsec:reconn}

\subsection{Spatial distribution and amplitude of HORNET}
\label{subsubsec:2DplotsRecon}

\begin{figure*}
    \includegraphics[width=6.8in]{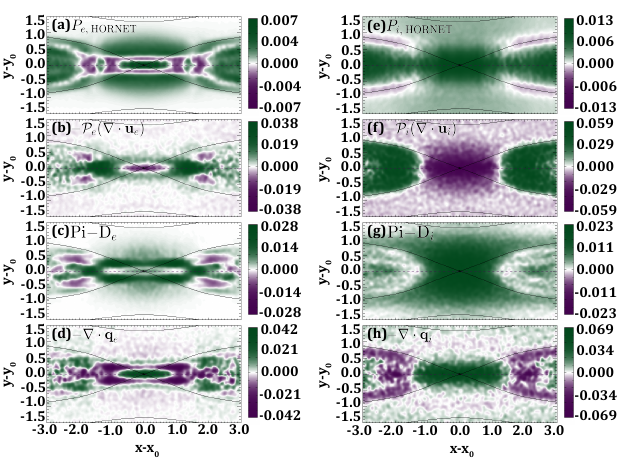}
    \caption{Particle-in-cell simulation results of power densities for electrons and ions during the \textcolor{black}{non-linear growth} phase of reconnection ($t = 13$). (a) and (e) higher order non-equilibrium terms \textcolor{black}{effective} power density $P_{e, {\rm HORNET}}$ and $P_{i, {\rm HORNET}}$, (b) and (f) pressure dilatation $-{\cal P}_e (\boldsymbol{\nabla}\cdot {\bf u}_e)$ and $-{\cal P}_i (\boldsymbol{\nabla}\cdot {\bf u}_i)$, (c) and (g) ${\rm Pi-D}_e$ and ${\rm Pi-D}_i$, and (d) and (h) \textcolor{black}{vector} heat flux density divergence $-\boldsymbol{\nabla}\cdot {\bf q}_e$ and $-\boldsymbol{\nabla} \cdot {\bf q}_i$.}
    \label{fig:Recondegreeoffreedomchanges}
\end{figure*}

Simulation results of $P_{\sigma, {\rm HORNET}}$ at $t = 13$ are shown in Fig.~\ref{fig:Recondegreeoffreedomchanges}, with electrons in (a) and ions in (e). Representative magnetic field line projections are in black, and we shift the coordinate system to be relative to the X-line location at $(x_0,y_0) = (9.59375,1.59375)$. The EDR is approximately located at $|x-x_0| \lesssim 2, |y-y_0| \lesssim  0.4$.

We begin with a discussion of the spatial structure of $P_{\sigma,{\rm HORNET}}$. For the electrons, we note that Ref.~\cite{Cassak_FirstLaw_2023} included a plot in their Fig.~3(e) of $d{\cal E}_{e,{\rm rel}}/dt$ for a reconnection simulation that only differed from the present simulation by its PPG. Since $P_{\sigma,{\rm HORNET}} = -n_\sigma d{\cal E}_{e,{\rm rel}}/dt$ from Eq.~(\ref{eq:pwrHORNET}), the structure of $P_{e,{\rm HORNET}}$ is largely similar to $-d{\cal E}_{e,{\rm rel}}/dt$ from Ref.~\cite{Cassak_FirstLaw_2023}. Consequently, we provide only an abbreviated analysis of the electrons here.

Upstream of the EDR, electrons become trapped in the bent magnetic field lines, leading to phase space densities that are elongated in velocity space in the direction parallel to the local magnetic field \cite{Egedal13}. 
\textcolor{black}{This implies the shape of the phase space density is becoming less Maxwellian} in the Lagrangian reference frame, which is associated with positive $P_{e,{\rm HORNET}} \simeq 0.003$. Near the X-line, triangular striated phase space densities develop due to electron Speiser orbits \cite{Speiser65,Ng_2011_PRL,shuster_spatiotemporal_2015}, a further evolution away from Maxwellianity, therefore associated with positive $P_{e,{\rm HORNET}} \simeq 0.007$. Interestingly, sandwiched between these two regions, there is a slim boundary extending from the two inflow regions ($0.2 < |y-y_0| < 0.3$) to the two outflow regions ($0.7 < |x-x_0| < 1.2$) inside the EDR in which $P_{e, {\rm HORNET}} \simeq -0.0015$ is weakly negative. Physically, this happens because the upstream phase space densities are elongated in the parallel direction \cite{Egedal13}, and positions close enough to the X-line have another population due to the Speiser orbits associated with motion perpendicular to the magnetic field. The resultant 
\textcolor{black}{phase space density} in the $(v_x,v_y)$-plane is \textcolor{black}{closer to being a} 
Maxwellian. 
Therefore, an electron fluid element entering the EDR has a negative $P_{e,{\rm HORNET}}$. Phase space densities in relevant regions are plotted in Appendix~\ref{sec:appendixe}. Thermalization of the outflowing electron jets downstream of the X-line \cite{Egedal12,Lavraud16,Wang16,Norgren20} is observed at the EDR outflow edge ($ 1.6 < |x-x_0| < 2.0$), associated with negative $P_{e,{\rm HORNET}} \simeq -0.002$.

For the ions, the ion diffusion region (IDR) is expected to have a vertical extent of $|y-y_0| < 2.24,$ but the simulation domain is too small to allow full ion coupling to the reconnection process; the simulation domain would have to be $\geq 40 \times 2.24 \simeq 90$ for the ions to fully couple \cite{Pyakurel19}. Upstream of the EDR, Fig.~\ref{fig:Recondegreeoffreedomchanges}(e) demonstrates that $P_{i,{\rm HORNET}}$ is weakly positive and gradually increases approaching the X-line, where it has a maximum value of $\simeq 0.013$. This indicates that ions become more non-Maxwellian (in the Lagrangian frame) approaching the X-line. This is consistent with expectations since ions demagnetize when their gyroradii exceed the distance from their guiding center to the magnetic field reversal, so only the most energetic ions are demagnetized further from the X-line and more ions are demagnetized closer to the X-line. 

The ions in the exhaust just downstream of the EDR have positive $P_{i,{\rm HORNET}}$ with characteristic value $0.006$, indicating an evolution away from Maxwellianity in these regions. This is likely from demagnetizing ions crossing the separatrix rather than crossing through the EDR. We expect for a larger system and at later times that ions would thermalize further downstream from the EDR, which would be associated with $P_{i,{\rm HORNET}} < 0$,
but we do not observe it in our simulation since the system is not in the steady-state and the simulation domain is small. Interestingly, starting at approximately $|x-x_0| \simeq 1.6$ along the separatrices outside of the EDR, $P_{i, {\rm HORNET}}$ is weakly negative with a value of $\simeq -0.001$. This implies the ion phase space density becomes more Maxwellian (in the Lagrangian frame) there. 

\subsection{Comparison to other power densities} 
\label{subsubsec:1DEDRlineplots}

We first perform a comparison of $P_{\sigma,{\rm HORNET}}$ with the power densities describing changes to the internal energy per particle described in Eq.~(\ref{eq:intenergyevolve}) at $t=13$ in order to determine the relative importance of each as a function of position during the \textcolor{black}{non-linear growth} phase. For electrons, the pressure dilatation $-{\cal P}_\sigma (\boldsymbol{\nabla} \cdot {\bf u}_\sigma)$ is shown in Fig.~\ref{fig:Recondegreeoffreedomchanges}(b), ${\rm Pi-D}_\sigma$ is shown in Fig.~\ref{fig:Recondegreeoffreedomchanges}(c), and  Fig.~\ref{fig:Recondegreeoffreedomchanges}(d) exhibits the negative divergence of the vector heat flux density $-\boldsymbol{\nabla} \cdot {\bf q}_\sigma$. These quantities are repeated in  Fig.~\ref{fig:Recondegreeoffreedomchanges}(f) - (h) for ions.

We first discuss electrons.  In a recent study with a simulation with the same initial conditions except with PPG = 25,600 instead of 6,400 \cite{Barbhuiya_PiD3_2022}, $-{\cal P}_e (\boldsymbol{\nabla} \cdot {\bf u}_e)$ and ${\rm Pi-D}_e$ were extensively discussed, but the \textcolor{black}{vector} heat flux density divergence was not analyzed. Near the X-line, we find $-\boldsymbol{\nabla} \cdot {\bf q}_e$ is positive (the \textcolor{black}{vector} heat flux is converging) with a value near 0.04 (panel (d)), which by itself would increase the electron internal energy. The pressure dilatation is negative with a value near $-0.015$ at the X-line (panel (b)), which by itself would decrease the electron internal energy.  ${\rm Pi-D}_e$ is weakly positive at 0.005 near the X-line (panel (c)). The net change to the internal energy is positive at this time, and the sum of the pressure dilatation, ${\rm Pi-D}_e$, and negative \textcolor{black}{vector} heat flux divergence is approximately 0.03. Concomitantly, panel (a) reveals that $P_{e,{\rm HORNET}}$ is positive at the X-line with a value near $0.007$. Thus, the electron phase space densities are becoming more non-Maxwellian with \textcolor{black}{an effective} power density approximately 23\% of that going to changing the electron internal energy. In contrast, near the outflow edges of the EDR ($|x-x_0| \simeq 1.7$), $-{\cal P}_e (\boldsymbol{\nabla} \cdot {\bf u}_e) \simeq 0.04$ and ${\rm Pi-D}_e \simeq 0.02$, whereas $-\boldsymbol{\nabla} \cdot {\bf q}_e \simeq -0.01$ with $P_{e,{\rm HORNET}} \simeq 0.002$. This indicates that the electron internal energy is increasing, while the electron phase space densities are becoming more Maxwellian at this location.

Looking at ions in the vicinity of the X-line, we observe a strongly negative pressure dilatation (panel (f)), reaching values of approximately $-0.06$ which by itself would decrease the ion internal energy through expansion near the X-line. In contrast, ${\rm Pi-D}_i$ is positive throughout the EDR and most of its surroundings, with a characteristic value near $0.02$. Near the X-line, we find $-\boldsymbol{\nabla} \cdot {\bf q}_i$ is positive (the vector heat flux density is converging) \cite{Eastwood_2020_PRL} with a value $\simeq 0.06$ (panel (h)), which by itself would increase the ion internal energy. Panel (e) shows that $P_{i,{\rm HORNET}}$ is positive in the vicinity of the X-line with a value of approximately 0.013. These results imply that, like the electrons, there is a local net increase near the X-line of the ion internal energy, and the sum of the three terms gives a net power of approximately 0.02, and the ion phase space density evolves away from Maxwellianity. For ions, the \textcolor{black}{effective} power going into changing the shape of the phase space density is approximately 65\% of the net power due to pressure dilatation, ${\rm Pi-D}_i$, and \textcolor{black}{vector} heat flux divergence near the X-line. Near the downstream edge of the EDR, pressure dilatation is instead positive with value near $0.05$, whereas ${\rm Pi-D}_i$ is negligible. We also see that $-\boldsymbol{\nabla} \cdot {\bf q}_i \simeq -0.06$, with $P_{i,{\rm HORNET}} \simeq 0.006$. Thus, contrary to what is observed for electrons, there is a local decrease in the ion internal energy and the ion phase space density is becoming more non-Maxwellian. 

We note that $-{\cal P}_i (\boldsymbol{\nabla} \cdot {\bf u}_i)$ and ${\rm Pi-D}_i$ were plotted in a previous simulation study \cite{Pezzi21} using a different code and system parameters (see their Fig.~1 for comparison). Their data was taken during the steady-state phase while ours is during the \textcolor{black}{non-linear growth} phase. We observe some differences: (i) in our simulation, $-{\cal P}_i (\boldsymbol{\nabla} \cdot {\bf u}_i)$ is notably more localized around the X-line in the $\hat{x}$ direction [see Fig.~\ref{fig:Recondegreeoffreedomchanges}(f)], presumably because our system size is smaller (theirs is 25 $\times 25$); and (ii) ${\rm Pi-D}_i$ has a single positive peak signature at the X-line in our simulation, while it has a double-peaked structure in Ref.~\cite{Pezzi21}, presumably because theirs is in the steady state. 


We also note the spatial distribution of $-\boldsymbol{\nabla} \cdot {\bf q}_\sigma$ for electrons and ions reveal structural similarities, except the ion structures are on larger scales.  
This is consistent with previous work that shows $-\boldsymbol{\nabla} \cdot {\bf q}_\sigma$ can be non-zero and dynamically significant locally \cite{Du_energy_2020_PRE, song_forcebalance_2020}, even though it vanishes when integrated over the whole periodic computational domain \cite{Yang_2022_ApJ}.
Physically, a local non-zero \textcolor{black}{vector} heat flux density emerges from velocity-space asymmetries of phase space densities. We provide an explanation of the \textcolor{black}{vector} heat flux density structure and its divergence in Appendix~\ref{sec:appendixq}.

\begin{figure}
    \includegraphics[width=3.4in]{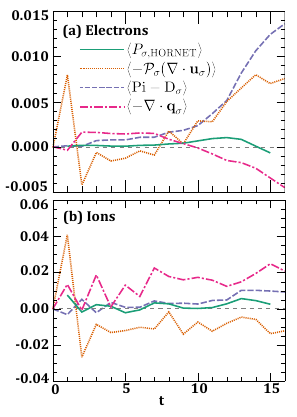}
    \caption{
    Instantaneous spatially-averaged power densities in the reconnection electron diffusion region as a function of time for (a) electrons and (b) ions. $\langle P_{\sigma,{\rm HORNET}} \rangle$ are the solid green lines, $\langle-{\cal P}_\sigma (\boldsymbol{\nabla} \cdot {\bf u}_\sigma)\rangle$ are the dotted orange lines, $\langle{\rm Pi-D}_\sigma\rangle$ are the dashed purple lines, and $\langle-\boldsymbol{\nabla} \cdot {\bf q}_\sigma\rangle$ are dash-dotted magenta lines. }
    \label{fig:ReconEDRboxavg}
\end{figure}

We now quantitatively compare the signed spatial averages of each of the previously discussed power densities, for electrons and ions separately, in order to compare their importance over a more extended region and to see how they evolve as a function of time in the simulation. We fix a box centered at the X-line of size $4 \ d_{i0}$ in length and $0.8 \ d_{i0}$ in width (representing the EDR size at $t=13$), and then calculate the signed spatial average (denoted by $\langle \rangle$) of each power density in this region as a function of time. The result is shown in Fig.~\ref{fig:ReconEDRboxavg}(a) and (b) for electrons and ions, respectively. In each, solid green denotes $\langle P_{\sigma,{\rm HORNET}} \rangle$, dotted orange denotes $\langle-{\cal P}_\sigma (\boldsymbol{\nabla} \cdot {\bf u}_\sigma)\rangle$, dashed purple denotes $\langle{\rm Pi-D}_\sigma\rangle$, and $\langle-\boldsymbol{\nabla} \cdot {\bf q}_\sigma\rangle$ is represented by dash-dotted magenta. 

At early time in both panels ($t \lesssim 3$), there is an initial positive spike of pressure dilatation, followed by an overshoot where it goes negative. This occurs because the initial conditions of the simulation contain a single drifting Maxwellian population, instead of two populations as is needed for the Harris sheet which is an exact kinetic equilibrium. The system self-consistently adjusts as the simulation is evolved. After this, most of the power densities are small until $t \simeq 12$ when the reconnection rate begins to increase rapidly in time.  Consequently, we focus mostly on these later times.

For the electrons, panel (a) reveals that $\langle {\rm Pi-D}_e \rangle$ and $\langle -{\cal P}_i (\boldsymbol{\nabla} \cdot {\bf u}_e) \rangle$ are the two largest terms, nearly equal early in the \textcolor{black}{non-linear growth} phase with $\langle{\rm Pi-D}_e \rangle$ becoming larger in the late \textcolor{black}{non-linear growth} phase. $\langle P_{e,{\rm HORNET}} \rangle$ remains relatively small throughout the evolution. We know from Fig.~\ref{fig:Recondegreeoffreedomchanges}(a) that $P_{e,{\rm HORNET}}$ is non-zero locally, so this result suggests that the net positive and negative \textcolor{black}{effective} power density contributions within the EDR associated with 
\textcolor{black}{evolution towards or away from LTE} mostly cancels out. Focusing on the net power densities at $t=13$ for electrons, we find $\langle{\rm Pi-D}_e\rangle \simeq 0.008$, $\langle-{\cal P}_e (\boldsymbol{\nabla} \cdot {\bf u}_e)\rangle \simeq 0.006$ and $\langle-\boldsymbol{\nabla} \cdot {\bf q}_e\rangle \simeq -0.002$, giving a net power density to change the electron internal energy of 0.012. Since $\langle P_{e,{\rm HORNET}}\rangle \simeq 0.001$, we find $\langle P_{e,{\rm HORNET}}\rangle \simeq 0.13 \ \langle{\rm Pi-D}_e\rangle \simeq 0.17 \ \langle-{\cal P}_e (\boldsymbol{\nabla} \cdot {\bf u}_e)\rangle$. \textcolor{black}{Because of the smallness of the average values and potential uncertainties in the measurements of $P_{\sigma,{\rm HORNET}}$, these fractions could have significant uncertainties.  The raw values indicate}  
that $\langle P_{e,{\rm HORNET}}\rangle$ 
represents a small but non-negligible 
\textcolor{black}{percentage} of about 8\% of the rate of change of energy density going into changing the electron internal energy.  


Interestingly, \textcolor{black}{Fig.~\ref{fig:ReconEDRboxavg} shows that} $\langle P_{e,{\rm HORNET}} \rangle$ becomes negative at $t=15$, \textcolor{black}{suggesting} that the net effect in the EDR is for the electron phase space density to evolve to be more Maxwellian at this time, in contrast to $t = 13$ where it becomes less Maxwellian. \textcolor{black}{While this could be impacted by the level of uncertainty in the measurements of $P_{e,{\rm HORNET}}$, there is a physical reason to believe this could be valid.}  As seen in Fig.~\ref{fig:Recondegreeoffreedomchanges}(a), the strongest negative $P_{e,{\rm HORNET}}$ regions are at the downstream edges of the EDR where electrons thermalize.
The electrons start to thermalize over a larger area as reconnection proceeds through its \textcolor{black}{non-linear growth} phase, so these patches of $P_{e,{\rm HORNET}} < 0$ grow with time (not shown), leading to the overall change in sign \textcolor{black}{of $\langle P_{e,{\rm HORNET}}\rangle$.}

For the ions shown in panel (b) the dominant positive energy conversion channel is $\langle-\boldsymbol{\nabla} \cdot {\bf q}_i\rangle$, which is nearly balanced by a negative $\langle -{\cal P}_i (\boldsymbol{\nabla} \cdot {\bf u}_i) \rangle$. There is a small net positive $\langle {\rm Pi-D}_i \rangle$ late in the \textcolor{black}{non-linear growth} phase ($t \geq 12$), that leads to a very weak increase in ion internal energy. The net \textcolor{black}{effective} energy conversion associated with \textcolor{black}{the evolution towards or away from LTE,} 
\textit{i.e.,} $\langle P_{i,{\rm HORNET}}\rangle$, though positive, is small compared to the other ion terms. This is expected from the 2D plot of $P_{i,{\rm HORNET}}$ (see Fig.~\ref{fig:Recondegreeoffreedomchanges}(e)) that shows it is non-negative locally inside the EDR leading to a net \textcolor{black}{evolution of ions towards LTE.} 
These results are very sensible physically, as the system size is not large enough for the ions to fully couple to the reconnection process in the simulation, so their net energy conversion is quite weak.
At $t=13$, we find $\langle-\boldsymbol{\nabla} \cdot {\bf q}_i\rangle \simeq 0.015$, $\langle{\rm Pi-D}_i\rangle \simeq 0.01$, and $\langle-{\cal P}_i (\boldsymbol{\nabla} \cdot {\bf u}_i)\rangle \simeq -0.005$, for a total of 0.02 going into ion internal energy. In comparison, $\langle P_{i,{\rm HORNET}}\rangle \simeq 0.007 \simeq 70 \% \ \langle{\rm Pi-D}_i\rangle \simeq 50 \% \ \langle -\boldsymbol{\nabla}\cdot {\bf q}_i\rangle$. Thus, in contrast to what we see for electrons, a larger 
\textcolor{black}{percentage} (35\%) of the energy going into ion internal energy goes to 
\textcolor{black}{the evolution of the ions towards LTE.} 

\textcolor{black}{We briefly examine the comparison between $P_{\sigma, {\rm HORNET}}$ and $\mathbf{J}_\sigma \cdot \mathbf{E}$. At $t=13$, we observe that the peak amplitude of $\mathbf{J}_\sigma \cdot \mathbf{E}$ exceeds that of the largest energy conversion metric (\textit{i.e.}, $-\boldsymbol{\nabla}\cdot {\bf q}_\sigma$) by more than a factor of two (not shown). Moreover, it surpasses $P_{\sigma,{\rm HORNET}}$ by more than an order of magnitude. This is expected because, in reconnection, a significant portion of the electromagnetic energy is converted into bulk kinetic energy of the particles, so the portion associated with internal energy and $P_{\sigma,{\rm HORNET}}$ are smaller.}  In summary, the results of this section suggest $P_{\sigma,{\rm HORNET}}$ is a useful tool for assessing the relative rates \textcolor{black}{at which the phase space density evolves towards or away from LTE and for comparing the rate to standard power densities} 
during the reconnection process.

\section{Results: Decaying Turbulence} \label{subsec:turbu}

\subsection{Spatial distribution and amplitude of HORNET}
\label{subsubsec:2DplotsTurb}

\begin{figure*}
    \includegraphics[width=6.8in]{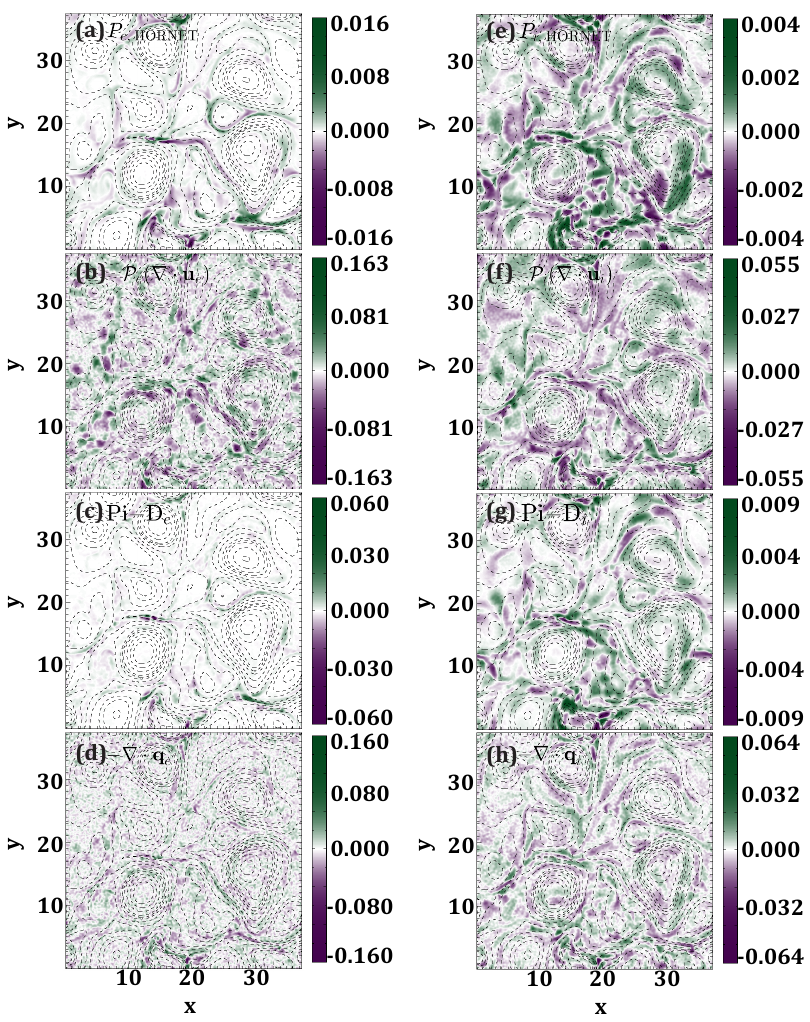}
    \caption{Particle-in-cell simulation results comparing power densities for electrons and ions during decaying turbulence at $t=30$. (a) and (e) higher order non-equilibrium terms \textcolor{black}{effective} power density $P_{e, {\rm HORNET}}$ and $P_{i, {\rm HORNET}}$, (b) and (f) pressure dilatation $-{\cal P}_e (\boldsymbol{\nabla} \cdot {\bf u}_e)$ and $-{\cal P}_i (\boldsymbol{\nabla} \cdot {\bf u}_i)$, (c) and (g)  ${\rm Pi-D}_e$ and ${\rm Pi-D}_i$, (d) and (h) \textcolor{black}{vector} heat flux density divergence $-\boldsymbol{\nabla} \cdot {\bf q}_e$ and $-\boldsymbol{\nabla}\cdot {\bf q}_i$.}
    \label{fig:Turbdegreeoffreedomchanges}
\end{figure*}

We now discuss the structure of $P_{\sigma,{\rm HORNET}}$ in the decaying turbulence simulation using 2D position-space plots of the various power densities in Fig.~\ref{fig:Turbdegreeoffreedomchanges} at $t=30 \sim 2 \ \tau_{nl}$, 
with representative in-plane magnetic field line projections as dashed-black lines.  This time is chosen as it is after the time at which mean square current density $\langle J^2 \rangle$ peaks (plot not shown), The panels are of the same quantities and are in the same order as those plotted in Fig.~\ref{fig:Recondegreeoffreedomchanges} for the reconnection simulation. 


We first consider electrons. In Fig.~\ref{fig:Turbdegreeoffreedomchanges}(a), we see that $P_{e, {\rm HORNET}}$ has substantially non-zero signatures co-located with intermittent structures (current sheets).  The regions with strongest $P_{e,{\rm HORNET}}$ up to a magnitude of about 0.016 are centered around $(x,y) \simeq (18, 36),~(21,2),~(29,5),~(15,18)$, $(15,4)$, and $(30,22)$. $P_{e,{\rm HORNET}}$ can have either sign in and around these structures.
We observe one location at $(x,y) \simeq (15, 7)$ around a current sheet where the structure of $P_{e,{\rm HORNET}}$ somewhat resembles that of the reconnection X-line shown in Fig.~\ref{fig:Recondegreeoffreedomchanges}(a), with a weak positive value surrounded by a negative value. We do not see this feature at other current sheets in our turbulence simulation. The absence is 
possibly due to the presence of asymmetries \cite{Servidio_PRL_2009} or flow shear arising at reconnection sites in the decaying turbulent system or the presence of the mean (guide) field in the turbulence simulation but not in the reconnection simulation that may modify the structure of $P_{e,{\rm HORNET}}$.  Interestingly, we observe numerous bands with non-zero $P_{e,{\rm HORNET}}$ coinciding with extended current sheets that are multiple ion-scales in length ($\sim 10 \ d_i$) and electron-scale in thickness. 

Panel (a) makes it appear that $P_{e,{\rm HORNET}}$ is small away from the current sheets, but this is an artifact of the colorbar because the largest values are significantly higher than values in the rest of the domain. In Appendix~\ref{sec:appendixeHORNET}, we reproduce the same plot but with the colorbar saturated at the maximum value used for $P_{i,{\rm HORNET}}$ to better observe the structures in $P_{e,{\rm HORNET}}$ away from the current sheets. 
This plot makes it clear that there are regions with non-zero of $P_{e,{\rm HORNET}}$ away from the current sheets. There are noticeable features in $P_{e,{\rm HORNET}}$ within ion-scale eddies of size 5-10 $d_{i}$. For example, in eddies centered around $(x,y) \simeq (12, 12),~(29, 16)$, and $(15,32)$, we see that the edges exhibit a weak but non-zero signature of $P_{e,{\rm HORNET}}$, while the centers of eddies have nearly exclusively negligible $P_{e,{\rm HORNET}}$.

Turning to the ions, Fig.~\ref{fig:Turbdegreeoffreedomchanges}(e) shows that almost the whole simulation domain has noticeable $P_{i, {\rm HORNET}}$, implying \textcolor{black}{the phase space densities evolve towards or away from LTE in regions away from the} 
intermittent structures. $P_{i,{\rm HORNET}}$ is found to be non-zero in the ion-scale eddies, \textit{e.g.,} at regions centered around $(x,y) \simeq (12, 12)$, $(30,27)$, and $(29,16)$. This makes sense physically, as ions are prone to get demagnetized in ion-scale magnetic islands since their characteristic ion gyroradius in the mean field is $0.77$.

The peak amplitude of $P_{e,{\rm HORNET}}$ is higher by a factor of $\simeq$ 4 than $P_{i,{\rm HORNET}}$ at this time in the simulation, 
\textcolor{black}{implying the peak effective power density of evolution towards or away from LTE is higher for electrons than ions.} Since the density and temperature for the ions and electrons are initially equal, Eq.~(\ref{eq:pwrHORNET}) suggests the electrons become more non-Maxwellian than the ions.

\subsection{Comparison to other power densities} 
\label{subsubsec:1DFullBoxlineplots}

We next compare $P_{\sigma,{\rm HORNET}}$ with power densities describing changes to internal energy.  
Pressure dilatation and ${\rm Pi-D}_\sigma$ have been the subject of many recent 2D turbulence simulation studies \cite{yang_PRE_2017,Yang17,Pezzi19,Pezzi21,Rueda_2022_ApJ}, with a greater emphasis on ${\rm Pi-D}_\sigma$. Here, we discuss the overall 2D structures in $-{\cal P}_\sigma (\boldsymbol{\nabla} \cdot {\bf u}_\sigma)$ and ${\rm Pi-D}_\sigma$ and compare their structures with $-\boldsymbol{\nabla} \cdot {\bf q}_\sigma$ and $P_{\sigma, {\rm HORNET}}$. 

We first analyze electron power densities at $t=30$, shown in Fig.~\ref{fig:Turbdegreeoffreedomchanges}.
We find that $-{\cal P}_e (\boldsymbol{\nabla} \cdot {\bf u}_e)$ (panel b)) has localized regions of compression or expansion on the order of 1~$d_i$ ($=5 \ d_e$) in size and up to an amplitude of $0.16$, unlike the structures seen in $P_{e, {\rm HORNET}}$. This is presumably due to electron plasma waves propagating through the system. Looking at ${\rm Pi-D}_e$ (panel (c)), we only see a few localized regions with prominent amplitudes up to an amplitude of about 0.06 that are a few ($\sim 5$) $d_{i}$ in length and electron-scale in thickness, which is similar to the structures seen in $P_{e, {\rm HORNET}}$. Finally, the 2D structures seen in $-\boldsymbol{\nabla} \cdot {\bf q}_e$ (panel (d)) show localized regions with significant amplitude up to $0.160$ that are co-located with current sheets, many of which coincide with regions of strong $P_{e, {\rm HORNET}}$ and ${\rm Pi-D}_e$. However, $-\boldsymbol{\nabla} \cdot {\bf q}_e$ is appreciably non-zero with positive and negative values throughout the simulation domain, with structure sizes on electron scales. Furthermore, at $(x,y) \simeq (15,7)$, the structure in $-\boldsymbol{\nabla} \cdot {\bf q}_e$ is not coherent while it is in the reconnection simulation, even though $P_{e, {\rm HORNET}}$ at the same location looks similar in the two. Amongst the four electron power densities, $-{\cal P}_e (\boldsymbol{\nabla} \cdot {\bf u}_e)$ and $-\boldsymbol{\nabla} \cdot {\bf q}_e$ are the terms with the largest local magnitudes, but there are more regions with the peak values in $-{\cal P}_e (\boldsymbol{\nabla} \cdot {\bf u}_e)$ than $-\boldsymbol{\nabla} \cdot {\bf q}_e$ as $-\boldsymbol{\nabla} \cdot {\bf q}_e$ is strongest where reconnection is taking place while $-{\cal P}_e (\boldsymbol{\nabla} \cdot {\bf u}_e)$ is strongest due to plasma waves pervasive in the domain.

We turn to ion power densities.
The 2D structures seen for $-{\cal P}_i (\boldsymbol{\nabla} \cdot {\bf u}_i)$ (panel (f)) and ${\rm Pi-D}_i$ (panel (g)) are of the order of a few $d_i$ size, similar to what is seen in $P_{i, {\rm HORNET}}$. Both have their highest magnitude near intermittent structures ($\simeq 0.055$ for pressure dilatation, $\simeq 0.009$ for ${\rm Pi-D}_i$), with smaller magnitudes elsewhere in the domain, and both quantities have various regions with positive and negative values. 
$-\boldsymbol{\nabla} \cdot {\bf q}_i$ (panel (h)) shows structures similar to the other power densities, with an amplitude up to $0.064$, with one notable difference that the structures are slightly smaller in size. Analogous to the electrons, the largest local magnitudes in ion power density are in $-\boldsymbol{\nabla} \cdot {\bf q}_i$ and $-{\cal P}_i (\boldsymbol{\nabla} \cdot {\bf u}_i)$, with the former having slightly higher peak magnitude.

\begin{figure}
    \includegraphics[width=3.4in]{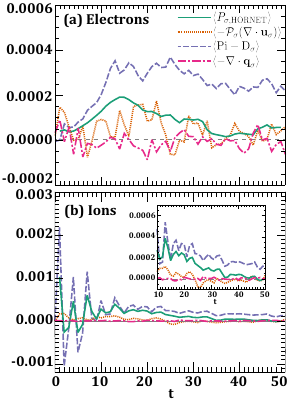}
    \caption{Instantaneous spatially-averaged power densities  in the turbulence simulation domain as a function of time for (a) electrons and (b) ions. The inset in (b) is a plot from time $t = 10 - 50$. The same color scheme as Fig.~\ref{fig:ReconEDRboxavg} is used. In this simulation, $ \tau_{nl} \simeq 17$.}
    \label{fig:Turbboxavg}
\end{figure}

Next, we compare the signed spatial average of the power densities considered here calculated over the whole simulation domain as a function of time in Fig.~\ref{fig:Turbboxavg}, with the same color scheme as in Fig.~\ref{fig:ReconEDRboxavg}. Panel (a) is for electrons, and panel (b) is for ions with the inset in the latter focusing from $t=10-50$. A similar analysis of $\langle-{\cal P}_\sigma (\boldsymbol{\nabla} \cdot {\bf u}_\sigma)\rangle$ and $\langle{\rm Pi-D}_\sigma\rangle$ was performed in Ref.~\cite{Yang_2022_ApJ} (see their Figs.~2(c) and (d)), which utilized a similar simulation as  used in the present study, except their simulation employed a lower PPG of 3,200, had a bigger simulation domain of $150^2 ~d_{i}^2$ and was evolved for more than $300~ \Omega^{-1}_{ci}$. We note that $\langle -\boldsymbol{\nabla} \cdot {\bf q}_\sigma\rangle \simeq 0$ for both species, as it should, since we are calculating the average over the entire computational domain.

We see in panel (b) that, up until about 1 $\tau_{nl}$, there are oscillations in $\langle{\rm Pi-D}_i\rangle$ and $\langle P_{i,{\rm HORNET}} \rangle$ that dwarf all other power densities. We attribute these oscillations 
to Alfv\'enic exchange between different modes from the initiation of the simulation, and therefore focus on times after $t \gtrsim 15 \simeq \tau_{nl}$.

For both electrons and ions, $\langle{\rm Pi-D}_\sigma\rangle$ is the dominant contribution to increasing the internal energy for this simulation, and there are oscillations around zero for $\langle-{\cal P}_e (\boldsymbol{\nabla} \cdot {\bf u}_e)\rangle$, consistent with Ref.~\cite{Yang_2022_ApJ}.  The value for $\langle{\rm Pi-D}_\sigma\rangle$ is comparable for the two species, starting at $\simeq 0.0003$ at $t \simeq 1 \ \tau_{nl}$ and slowly decreasing throughout the rest of the evolution as the turbulence decays.  We find $\langle P_{\sigma,{\rm HORNET}}\rangle$ is $\simeq 0.0002$ at $t \simeq \tau_{nl}$, \textcolor{black}{which} 
is $\simeq 67\%$ of $\langle {\rm Pi-D}_\sigma \rangle$ at that time. 
\textcolor{black}{While acknowledging that the smallness of the averages increases the potential impact of uncertainties, our turbulence simulation suggests
the effective power density associated with the phase space density evolving towards or away from LTE} 
is a significant 
\textcolor{black}{percentage} of the overall energy budget at this time.

We observe that $\langle P_{\sigma,{\rm HORNET}}\rangle$ decays in time faster than $\langle{\rm Pi-D}_\sigma\rangle$. By $t=30 \simeq 2 \ \tau_{nl}$, we find that $\langle P_{e,{\rm HORNET}}\rangle \simeq 31\%$ $\langle{\rm Pi-D}_e\rangle$ (down from 67\% at $t \simeq 1 \ \tau_{nl}$) and $\langle P_{i,{\rm HORNET}}\rangle \simeq 50\%$ $\langle{\rm Pi-D}_i\rangle$ (down from 67\% at $t \simeq 1 \ \tau_{nl}$). Physically, this implies that the plasma thermalizes more rapidly than it cools as the turbulence decays, whether through physical or numerical effects.

Interestingly, we find that $\langle P_{e,{\rm HORNET}}\rangle$ peaks at $t=14$, earlier than the mean square current density $\langle J^2\rangle$ (dominated by electrons) which peaks at $t=21$ (plots not shown). 
At $t = 14$, the domain has a single intermittent structure with strongly positive $P_{e,{\rm HORNET}}$, with its spatial distribution more closely resembling that seen in reconnection simulation in Fig.~\ref{fig:Recondegreeoffreedomchanges}(a) than the example discussed at $t=30$ in Fig.~\ref{fig:Turbdegreeoffreedomchanges}(a). 
Consequently, the peak 
\textcolor{black}{effective power density associated with electrons evolving away from LTE} occurs prior to current sheets reaching their maximum current as eddies collide. We attribute this to the fact that electron phase space densities start to become non-Maxwellian before electrons demagnetize in the current sheets due to electron trapping in mirror magnetic fields present near the current sheets \cite{Egedal13}.

\begin{figure*}
    \begin{center}
    \includegraphics[width=\textwidth]{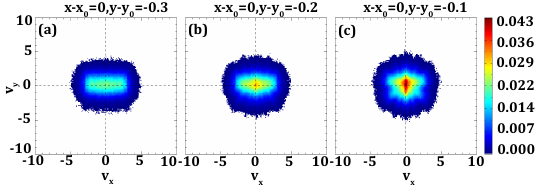}
    \caption{Reduced electron phase space densities $f_e(v_x,v_y)$ from a particle-in-cell magnetic reconnection simulation (using the simulation with PPG = 25,600 from Ref.~\cite{Barbhuiya_PiD3_2022}). (a) Inside the EDR, close to its upstream edge, 
    (b) further inside the EDR, 
    and (c) near the X-line.}
    \label{fig:UpstreamrEVDFs}
    \end{center}
\end{figure*}

\textcolor{black}{Finally, we briefly compare $P_{\sigma, {\rm HORNET}}$ and $\mathbf{J}_\sigma \cdot \mathbf{E}$.
in the decaying turbulence simulation at $t=30$. As with the pattern seen in $-\mathcal{P}_e (\boldsymbol{\nabla} \cdot \mathbf{u}_e)$ in Fig.~\ref{fig:Turbdegreeoffreedomchanges}, ${\bf J}_e \cdot {\bf E}$ exhibits localized regions of positive and negative amplitudes (not shown). We posit that the structure
may be attributed to propagating electron plasma waves within the simulation domain. This hypothesis is supported by the proximity of the peak amplitude of ${\bf J}_e \cdot {\bf E}$ (0.157) to that of $-\mathcal{P}_e (\boldsymbol{\nabla} \cdot \mathbf{u}_e)$, which is an order of magnitude higher than $P_{e,{\rm HORNET}}$. For the ions, 
the structures in ${\bf J}_i \cdot {\bf E}$ bear resemblance to those observed in $-\mathcal{P}_i (\boldsymbol{\nabla} \cdot \mathbf{u}_i)$ and ${\rm Pi-D}_i$, with peak amplitudes approximately two times larger than the other channels of energy conversion illustrated in Fig.~\ref{fig:Turbdegreeoffreedomchanges} (not shown). This higher value of ${\bf J}_\sigma \cdot {\bf E}$ occurs because it quantifies the rate of energy conversion from the fields to the particles, while the power densities considered here contain information about changes to internal energy but not bulk kinetic energy.}

\section{Discussion and Conclusion} \label{sec:conclusion}

In this work, we define the ``higher order non-equilibrium term" (HORNET) \textcolor{black}{effective} power density $P_{\sigma,{\rm HORNET}}$ [see Eq.~(\ref{eq:pwrHORNET})]. It is local in space and time and quantifies the rate at which \textcolor{black}{a species evolves towards or away from local thermodynamic equilibrium.}
Positive $P_{\sigma,{\rm HORNET}}$ implies the phase space density is locally becoming less Maxwellian in time, while negative $P_{\sigma,{\rm HORNET}}$ implies it is locally becoming more Maxwellian (thermalizing). The motivation for defining $P_{\sigma,{\rm HORNET}}$ is that it is a\textcolor{black}{n effective} power density, on the same footing as well-studied power densities describing changes to internal energy, namely pressure dilatation $-{\cal P}_\sigma (\boldsymbol{\nabla} \cdot {\bf u}_\sigma)$ describing compression, ${\rm Pi-D}_\sigma$ describing incompressible effects, and the divergence of the \textcolor{black}{vector} heat flux density $-\boldsymbol{\nabla} \cdot {\bf q}_\sigma$ [see Eq.~(\ref{eq:intenergyevolve})], as well as ${\bf J}_\sigma \cdot {\bf E}$. This allows the terms to be compared to determine the relative importance of \textcolor{black}{HORNET compared to the 
power} 
densities.

To exemplify the utility of $P_{\sigma,{\rm HORNET}}$, we use numerical simulations of plasmas that are not in LTE to study its spatial and temporal variation for both electrons and ions. We use fully kinetic PIC simulations of magnetic reconnection at the time when the reconnection rate is most rapidly increasing. We first look at the spatial variation of $P_{\sigma,{\rm HORNET}}$ in and around the EDR. We find that $P_{\sigma,{\rm HORNET}}$ identifies regions where particle phase space densities locally evolve toward or away from Maxwellianity in the comoving (Lagrangian) reference frame. We also find that $P_{\sigma,{\rm HORNET}}$ is locally significant at and near the X-line for both electrons and ions in our simulation. When compared to power densities describing changes to the internal energy, we find that $P_{e,{\rm HORNET}} \simeq 23\%$ of the sum of $-{\cal P}_e (\boldsymbol{\nabla} \cdot {\bf u}_e)$, ${\rm Pi-D}_e$, and $-\boldsymbol{\nabla} \cdot {\bf q}_e$, and $P_{i,{\rm HORNET}} \simeq 65 \%$ of the sum of $-{\cal P}_i (\boldsymbol{\nabla} \cdot {\bf u}_i)$, ${\rm Pi-D}_i$, and $-\boldsymbol{\nabla} \cdot {\bf q}_i$. We further examine the spatial average of $P_{\sigma,{\rm HORNET}}$ inside the EDR and compare it with the other power densities as a function of time. Though smaller in magnitude than the other power densities, $\langle P_{\sigma,{\rm HORNET}} \rangle$ represents  \textcolor{black}{the dynamical importance of the electrons departing from LTE. }
During the \textcolor{black}{non-linear growth} phase of reconnection (at $t=13$), $\langle P_{e,{\rm HORNET}}\rangle \simeq 8.3 \%$ of the sum of $\langle-{\cal P}_e (\boldsymbol{\nabla} \cdot {\bf u}_e)\rangle$, $\langle{\rm Pi-D}_e\rangle$ and $\langle-\boldsymbol{\nabla}\cdot {\bf q}_e\rangle$ and $\langle P_{i,{\rm HORNET}}\rangle \simeq 35 \%$ of the sum of $\langle-{\cal P}_i (\boldsymbol{\nabla} \cdot {\bf u}_i)\rangle$, $\langle{\rm Pi-D}_i\rangle$ and $\langle-\boldsymbol{\nabla}\cdot {\bf q}_i\rangle$.

We also study PIC simulations of decaying turbulence. We investigate the 2D spatial distribution of $P_{\sigma, {\rm HORNET}}$ and compare it with other power density measures. 
Quantitatively, we find $P_{\sigma,{\rm HORNET}}$ is comparable to other power densities. Similar to the reconnection simulation, we look at the time evolution of the spatial average of these energy conversion metrics, but averaged over the whole simulation domain. We find $\langle P_{\sigma,{\rm HORNET}} \rangle$ is a substantial fraction of the only other significant power density 
$\langle {\rm Pi-D}_\sigma \rangle$; at one non-linear time into the evolution of the system it is 67\% for both electrons and for ions.

\textcolor{black}{It is tempting to directly compare the prevalence of non-LTE effects in the reconnection and decaying turbulence simulations in this study. However, we carry out spatial averaging for $P_{\sigma,{\rm HORNET}}$ in the reconnection simulation over a localized domain, approximately the EDR, but for the whole domain in the decaying turbulence simulation, so the results for the percentage of $P_{\sigma,{\rm HORNET}}$ compared to the power densities are not describing the same physical quantity. Moreover, we find that the signed average of $P_{\sigma,{\rm HORNET}}$ is lower in the decaying turbulence simulation than in the reconnection simulation. This difference can be attributed to the fact that the regions undergoing reconnection, embedded within the turbulence simulation, are relatively small compared to the overall simulation domain. Furthermore, the turbulence simulation contains regions with both positive and negative $P_{\sigma, {\rm HORNET}}$, and these oppose each other when averaged over the entire domain. Thus, it is essential to exercise caution when comparing quantities across systems.}

We now address the implications of the present work. We discuss how one might expect the magnitudes of electron and ion HORNET \textcolor{black}{effective} power density to compare to each other. Using Eq.~(\ref{eq:pwrHORNET}), their ratio is
\begin{equation}
    \frac{P_{e,{\rm HORNET}}}{P_{i,{\rm HORNET}}} = \frac{n_e \mathcal{T}_e \frac{d}{dt} \left(s_{e v,{\rm rel}}/n_e\right)}{n_i \mathcal{T}_i \frac{d}{dt} \left(s_{i v,{\rm rel}}/n_i\right)}.
\label{eq:avgHORNETratio}
\end{equation}
For a singly ionized two-component quasi-neutral plasma, the densities cancel. Thus, the electron-to-ion HORNET ratio is the electron-to-ion temperature ratio times the ratio of the rate of change of the relative entropy per particle. Calculating this ratio provides information about which phase space density is more rapidly changing its shape. 
For example, averaging over the EDR at $t=13$ in the reconnection simulation, we find $\langle P_{e,{\rm HORNET}} \rangle/\langle P_{i,{\rm HORNET}} \rangle \simeq 0.14$. Since the initial temperature ratio $T_e/T_i = 0.2$, we conclude the two species are non-Maxwellianizing at a comparable rate, with the ions being at a slightly faster rate. 
Similarly, averaging over the whole domain at $t=30$ in the turbulence simulation, $\langle P_{e,{\rm HORNET}} \rangle/\langle P_{i,{\rm HORNET}} \rangle \simeq 1.1$. Since the initial temperature ratio is 1, we conclude the two species are non-Maxwellianizing at a comparable rate, with the electrons being slightly faster.

A second important implication applies for collisionless or weakly collisional plasmas, such as some in space. If the collisional terms in Eq.~(\ref{eq:dsvrelondt}) are negligible, then $P_{\sigma,{\rm HORNET}}$ is equivalent to
\begin{equation}
    P_{\sigma,{\rm HORNET}} = {\cal T}_\sigma (\boldsymbol{\nabla} \cdot {\bf J}_{\sigma,{\rm th}})_{{\rm rel}}.
    \label{eq:HORNETnocol}
\end{equation}
This form could be useful because it is easier to calculate in simulations, and potentially in spacecraft or laboratory experiments, because it does not require taking a temporal derivative.

\begin{figure*}
    \begin{center}
    \includegraphics[width=\textwidth]{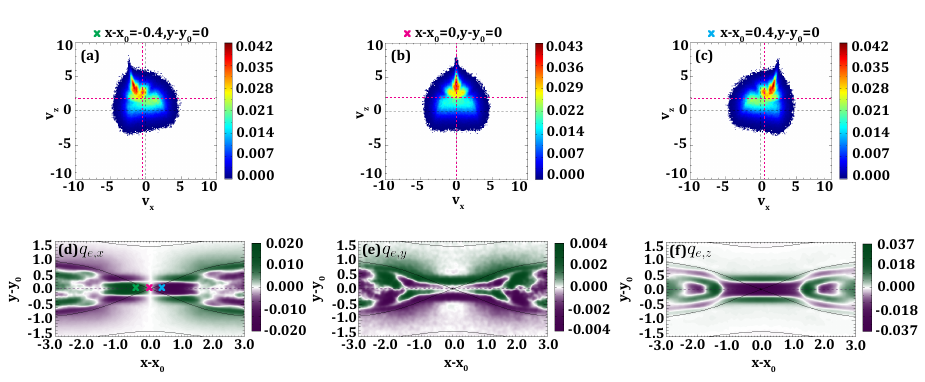}
    \caption{Particle-in-cell magnetic reconnection simulation results (using the simulation with PPG = 25,600 from Ref.~\cite{Barbhuiya_PiD3_2022}) to explain the vector heat flux density profiles observed. Reduced electron phase space densities $f_e(v_x,v_z)$ at the (a) green, (b) magenta, and (c) cyan ``x'''s overlayed on the $x$-component of electron heat flux density $\textbf{q}_e$ shown in panel (d). Panels (e) and (f) show the $y$ and $z$-components of  $\textbf{q}_e$. The magenta dashed lines in (a) - (c) denote the bulk flow velocity components $u_{e,x}$ and $u_{e,z}$.}
    \label{fig:HeatfluxandrEVDFs}
    \end{center}
\end{figure*}

We now discuss possible applications of this study.  In turbulent plasmas, energy injected at large scales cascades down to smaller scales \cite{matthaeus2021turbulence}, leading to intermittent coherent structures \cite{wan_intermittent_2012}.  It has been argued \cite{Yang17,yang_PRE_2017,Matthaeus20} that all channels of energy conversion are present in these structures, which convert between bulk flow, thermal and electromagnetic energies. The present results provide an approach to quantify the relative importance of \textcolor{black}{non-LTE effects,}
which we have shown can be a significant 
contribution. 
Magnetic reconnection in thin current sheets, both within turbulent plasmas \cite{matthaeus&lamkin_Turbulence_1986,Servidio_PRL_2009,Servidio_PoP_2010,haggerty_simulation_2017,retino_observation_2007,sundkvist_dissipation_2007,Phan18,Stawarz19} and in other solar, magnetospheric, planetary, astrophysical, and laboratory settings \cite{Zweibel09}, have long been investigated as locations of energy conversion. Understanding the contribution of non-LTE physics 
near the diffusion region and beyond is important to understand this conversion. 
Collisionless shocks, such as Earth's bow shock, convert bulk flow energy into magnetic energy and thermal and non-thermal energy at a relatively sharp boundary layer \cite{Marcowith16}. Plasmas in and around the boundary layer can be very strongly non-Maxwellian, so quantifying the contribution 
is expected to be very useful. Beyond plasma physics, we expect potential applications in sciences where non-LTE systems are routinely studied such as condensed matter physics \cite{Jaeger10}, open quantum systems \cite{Schaller14}, and molecular dynamics in chemistry and biology \cite{Evans86,Karplus90}.

We expect $P_{\sigma,{\rm HORNET}}$ to be useful in a variety of analysis techniques for non-LTE systems for which $f_\sigma$ can be measured such as numerical simulations, laboratory experiments, and spacecraft observations of space plasma. In particular, phase space densities are readily obtainable in particle-in-cell and Vlasov/Boltzmann simulations, thus calculating $P_{\sigma,{\rm HORNET}}$ is possible. Slices of phase space densities can be measured in plasma experiments \cite{Raymond_2018_PRE,Milder_2021_PoP,Milder_2021_PRL,Shi_2022_PoP,Shi_2022_PRL}, although full three-dimensional measurements are currently beyond the present technology. For spacecraft data, it has been shown \cite{Argall22,lindberg_2022_entropy,agapitov_2023_entropy} that kinetic entropy can reliably be measured using NASA's Magnetospheric Multiscale (MMS) satellites \cite{Burch16}. One direct avenue of future work is to measure $P_{\sigma,{\rm HORNET}}$ using MMS data during turbulence, reconnection, and 
collisionless shocks.

Another avenue for future work is to ascertain the impact of the system size on the 2D structure of $P_{i,{\rm HORNET}}$ during magnetic reconnection.  A limitation of the present reconnection study is the small system size; larger simulations are necessary to determine how $P_{i,{\rm HORNET}}$ will change in a system in which ions fully couple to large-scale physics \cite{Pyakurel19}.  In a simulation with a larger domain, we expect ions to recouple to the reconnected field near the downstream edges of the IDR and thermalize, which will appear as a $P_{i,{\rm HORNET}}<0$ signature, analogous to the negative $P_{e,{\rm HORNET}}$ seen for electrons near the downstream edges of the EDR in the present study. 
Other avenues for future work, for both reconnection and turbulence, include utilizing a more realistic mass ratio, performing 3D simulations, studying the parametric dependence on $P_{\sigma,{\rm HORNET}}$ with ambient plasma parameters, and studying the dependence on out-of-plane mean (guide) magnetic field, upstream asymmetries and/or upstream bulk flow. 

In the absence of physical collisions as in the PIC simulations discussed here, energy and entropy \textcolor{black}{are conserved reasonably well but} are not conserved \textcolor{black}{perfectly} due to numerical dissipation. This implies that the physical energy conversion to internal energy by the discussed energy conversion measures, \textit{i.e.,} pressure dilatation, $\mathrm{{Pi-D}}_\sigma$, and the divergence of the \textcolor{black}{vector} heat flux density, is not equal to the rate of change of internal energy density (in the Lagrangian frame) as suggested by Eq.~(\ref{eq:IntEnerEq}). Similarly, the rate of change in relative energy in Eq.~(\ref{eq:dsvrelondt2}) is affected by numerical dissipation. \textcolor{black}{Entropy metrics can be useful for quantifying numerical dissipation ({\it e.g.,} \cite{Liang20b}).} The effects of physical collisions and \textcolor{black}{entropic approaches to quantifying} numerical dissipation will be topics of future study.

\begin{acknowledgments}
P.A.C.~gratefully acknowledges primary support from NSF Grant PHY-1804428 and supplemental support from NSF Grant AGS-1602769, NASA Grant 80NSSC19M0146 and DOE Grant DE-SC0020294.  
M.R.A.~acknowledges support by NASA grant 80NSSC23K0409. This research used resources of the National Energy Research Scientific Computing Center (NERSC), a U.S. Department of Energy Office of Science User Facility located at Lawrence Berkeley National Laboratory, operated under Contract No. DE-AC02-05CH11231 using NERSC award FES-ERCAP0027083. Data used for the figures are available publicly at \url{https://doi.org/10.5281/zenodo.8147803}. We acknowledge basing our color table for 2D plots on the ``bam" color table \cite{Crameri_2020_SciColorMaps} in an effort to improve color-vision deficiency accessibility \cite{crameri_2020_CVD_NatCom, Brewer_2020_ColorMaps}. The colorbar extrema are green [$(r_g,g_g,b_g) = (2,73,29) = {\cal C}_g$] and purple [$(r_p,g_p,b_p) = (68,3,79) = {\cal C}_p$], where $(r,g,b)$ are the red, green, and blue components of each color, given by the subscript $g$ for green and $p$ for purple.  Bam varies linearly from purple to white to green, but we stretch the color table to provide higher contrast for small values. Letting ${\cal C}(k) = (r(k),g(k),b(k))$ be the color table with index $k$ from 0 to 255, we use a color table of ${\cal C}(k) = 255 - (255-{\cal C}_p) F(k)$ from $k = 0$ to 127 (purple) and ${\cal C}(k) = 255 - (255-{\cal C}_g) F(-k)$ from $k = 127$ to 254 (green), with $r(255) = g(255) = b(255) = 0$ (black) and $F(k)=\tanh{\left[3(127-k)/127\right]}/\tanh{3}$.

\end{acknowledgments}

\appendix
\section{Why $P_{e,{\rm HORNET}}$ is negative inside the EDR near its upstream edge during reconnection \textcolor{black}{non-linear growth phase}} 
\label{sec:appendixe}

In Sec.~\ref{subsubsec:2DplotsRecon}, we provided a physical reason for the negative $P_{e,{\rm HORNET}}$ inside the EDR between the upstream edge and the region near the X-line ($0.15 \lesssim |y-y_0| \lesssim 0.25$) during the \textcolor{black}{non-linear growth} phase of anti-parallel magnetic reconnection. Here, we provide further details. We show reduced electron phase space densities $f_e(v_x,v_y)$ (with $v_z$ integrated out) in Fig.~\ref{fig:UpstreamrEVDFs} at three locations from the reconnection simulation from Ref.~\cite{Barbhuiya_PiD3_2022} that is identical to the simulation presented in this study except PPG = 25,600. Panel (a) is at $(x-x_0,y-y_0)=(0,-0.3)$ inside the EDR, close to its upstream edge.  Here, $f_e$ is elongated in the parallel ($\approx \hat{x}$) direction, as explained by Egedal and colleagues \cite{Egedal13}. Panel (b) is at $(x-x_0,y-y_0)=(0,-0.2),$ closer to the X-line inside the EDR where $P_{e,{\rm HORNET}} < 0$.  Panel (c) is at $(x-x_0,y-y_0)=(0,-0.1),$ even closer to the X-line where electrons undergo Speiser orbits.  Panel (b) reveals that $f_e$ contains particles from the distributions in both (a) and (c), which reduces the anisotropy in $f_e$ going from panel (a) to panel (b).  Therefore, in the comoving reference frame for a plasma parcel convecting from outside the EDR to inside the EDR close to its upstream edge, the phase space density becomes more Maxwellian, so $P_{e,{\rm HORNET}}$ is negative, in agreement with the simulations.

\section{Physical Mechanism for the Vector Heat Flux Density Structure in the Reconnection Diffusion Region}
\label{sec:appendixq}

\begin{figure}
    \begin{center}
    \includegraphics[width=3.4in]{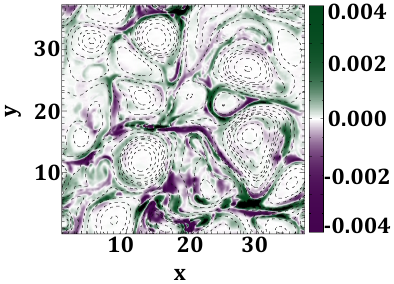}
    \caption{Plot of $P_{e, {\rm HORNET}}$ identical to Fig.~\ref{fig:Turbdegreeoffreedomchanges}(a), except with colorbar saturated at values identical to that of $P_{i, {\rm HORNET}}$ in Fig.~\ref{fig:Turbdegreeoffreedomchanges}(e).}
    \label{fig:TurbElectronHORNET_saturated}
    \end{center}
\end{figure}


At the X-line, the reduced phase space density is symmetric around $v_x =0$, so $q_{e,x} = 0$ (panel (d)). Similarly, $f_e$ is symmetric in $v_y$ (not shown), so $q_{e,y} = 0$ (panel (e)). However, the striated phase space density $f_e$ at the X-line is not symmetric in $v_z$. Relative to the bulk flow velocity $u_{e,z}$, the phase space density weighted by $v^{\prime 2}$ is more significant at $v_z < u_{e,z}$ than $v_z > u_{e,z}$, so there is a negative heat flux density $q_{e,z}$ at the X-line (panel (f)). As one goes away from the X-line in either outflow direction, the phase space density is rotated by the reconnected magnetic field $B_y$ along the $y-y_0=0$ line, as has been discussed previously \cite{Ng_2011_PRL,shuster_spatiotemporal_2015,Cassak_PiD1_2022,Egedal_2023_PoP}. To the left of the X-line, there is a negative $u_{e,x}$ (panel (a)). The part of the phase space density below the bulk velocity $u_{e,z}$ rotates in the positive $v_x$ direction, so this rotation creates an asymmetry in $f_e$ that gives a positive $q_{e,x}$, consistent with panel (d). Similarly, to the right of the X-line, $q_{e,x}$ is negative, also consistent with panel (d). 
This is the kinetic manifestation of the spatial structure in $q_{e,x}$. This term is the dominant contribution to the converging \textcolor{black}{vector} heat flux $-\boldsymbol{\nabla} \cdot {\bf q}_e$ at and near the X-line as seen by the green color near the X-line in Fig.~\ref{fig:Recondegreeoffreedomchanges}(d). 

\section{$P_{e,{\rm HORNET}}$ in the turbulence simulation}
\label{sec:appendixeHORNET}

In Sec.~\ref{subsubsec:2DplotsTurb}, we display a 2D plot of $P_{e,{\rm HORNET}}$ in a PIC simulation of decaying turbulence at $t = 30$ and discuss its structure.  To highlight the structure beyond the regions of its peak value, we show the same quantity shown in Fig.~\ref{fig:Turbdegreeoffreedomchanges}(a) in Fig.~\ref{fig:TurbElectronHORNET_saturated}, but with the colorbar saturated to have the same range as that for $P_{i,{\rm HORNET}}$ in Fig.~\ref{fig:Turbdegreeoffreedomchanges}(e). The regions with non-zero $P_{e,{\rm HORNET}}$ outside the intermittent structures are more clearly visible here.

\bibliography{HORNET}

\end{document}